# Privatizing user credential information of Web services in a shared user environment

Pinaki Mitra
Indian Institute of Technology Guwahati
pinaki@iitg.ernet.in

Rinku Das
Indian Institute of Technology Guwahati
rinku@iitg.ernet.in

Girish Sundaram
Technical Solution Architect
IBM India Software Labs
gisundar@in.ibm.com

*ABSTRACT*--User credentials security is one of the most important tasks in Web World. Most Web sites on the Internet that support user accounts store the users' credentials in a database. Now a days, most of the web browsers offer auto login feature for the favourite web sites such as yahoo, google, gmail etc. using these credential information. This facilitates the misuse of user credentials. Privatizing user credential information of web services in a shared user environment provides a feature enhancement where the root user will be able to privatize his stored credentials by enforcing some masking techniques such that even a user logs on to the system with root user credentials, he will not be able to access privatized data. In case of web browsers' auto login feature, a root user can disable the feature manually by deleting entries from web browsers' saved password list. But this involves spending a considerable amount of time and the biggest problem is that he has to insert those credentials once again when he next visits these websites. This application resumes auto login feature whenever root user disable the masked mode.

The application includes two parts: Masked Application Mode and Disabling the Masked Application Mode. When the system goes for masked application mode, the other user will not be able to use the credentials of the root user. If the other user tries to access any of the web pages which have been masked, the other user will have to authenticate with his own credentials. Disabling the masked mode requires authentication from the root user. As long as this credential is not shared, masked mode can be disabled only by the root user.

## I. INTRODUCTION

Many times it becomes essential for a user to share their physical machines with multiple users for various purposes. In such cases the user needs to share his root credentials with others. Then he needs to either delete all private data and to have a back up or create an third party user profile and limiting access capability. Sometimes it may not be possible since the other party would require root credentials to perform activities like upgrading a system or updating applications etc. Also now a days many applications store user credentials at OS service level and use them to start the application automatically. Hence we need a feature to privatize user credentials.

The system will have a new mode of operation called as "**Masked Application Mode**". The root user of the system enables this mode by executing a command at the command line. Once the command is invoked, a list of all web sites where root user's credentials are stored comes up. Root user then select the web sites which he wants to mask. Then an application masked those web sites and the currently logged user even if he is the root user will not be able to use auto login property. He has to provide his credentials explicitly. Now the root user can pass on the system control to other users. If the other user tries to access any masked web sites, he will be prompted for his credentials. The other user will have to authenticate with his own credentials. If only the other user has such privileges on web sites, he will be able to access these web sites.

Using the above explained idea we are able to ensure that even if the other user has the root credentials, he will not be able to misuse or access masked information.

To disable the masked mode, the original root user will need to authenticate using his personal credential. Once he authenticates, the system comes back to the normal mode of operation and all web sites will start using the cached root user credentials.

## II. RELATED APPROACHES

One approach to disable auto login to web sites is to delete saved passwords from a list of all the usernames and passwords that have been saved by the browser. This list is stored in browser toolbar. The "**Saved Passwords**" window (Firefox 3.00+) or "**Manage saved passwords**"(Google chrome) or "**Autocomplete**" window (IE) lists the web sites and user names for stored passwords. Firefox and Google chrome includes a "**Show Passwords**" button that lets us view stored passwords. Select the items that to delete and click the "**Remove**" button (Firefox) or click the **X** (Google chrome) that appears on the right. In Internet Explorer, from the Tools menu, select Internet Options. On the General tab, under "**Browsing history**", click **Delete...** One of the several options is **delete password**.

Another approach is deleting cookies. Automatic logins are made possible by storing cookies onto users' computer, so one solution would be to get rid of those cookies.

But main problem with these approaches is that user needs to insert credentials once again when he next visits these websites. Now-a-days one user has multiple accounts in single website. Hence these approaches can become tedious and time consuming.

## III. PROPOSED APPROACH

Following algorithm can be used as a high level implementing idea:

1. Root user enables **"Masked Application Mode".**
2. List the web sites which have stored user credentials.
3. Host user selects the web sites which he wants to **"mask".**
4. Store the selection list from Step 3 for future reference.
5. Prompt the user to log off and login again.
6. The system is now in masked mode.
7. Does the currently logged user try to use masked websites? If **Yes** then prompt for user authentication. If **No** then goto next step.
8. Does the currently logged user select to disable the masked mode? If **Yes** then prompt for currently logged user to use personalized authentication mechanism. If successful then disable **"Masked Application Mode".**
9. End.

For example **Firefox password manager** uses the file *signons.sqlite* to store the encrypted names and passwords which is a **SQLite3** database file. It also uses *key3.db* file to store the encryption key and that file need to match the signons file. The database *key3.db* contains a key used to encrypt and decrypt saved passwords. **SQLite** is a software library that implements a self-contained, serverless, zero configuration, transactional SQL relational database management system (**RDBMS**). Since *signons.sqlite* file is a database file, we can edit the file using standard SQL queries. The new *signons.sqlite* database file has two tables, **moz_disabledHosts** and **moz_logins**. The **moz_disabledHosts** table contains list of excluded websites which are exempted from storing passwords by user. The **moz_logins** table contains all the saved website passwords.

It is clear that we need to use the table called **moz_logins**. We can use standard sql query to extract the selected rows with entire entry and save it to another database table or file for future reference (As mentioned in implementing idea steps 3 and 4). The approach will be that first we select all rows from the table **moz_logins**. The result will be used to populate a list of web sites which has stored user credentials. User will select the web sites which he wants to mask. This selection will be used to execute another sql query to delete selected rows and saving the entries in another place. When user wants to disable the masked mode another sql query will be executed to reinsert those rows with previously saved credentials. Corresponding to each entry in table **moz_logins** there is an entry in *key3.db* file which stores encryption and decryption key for the particular entry. Deleting entries from table **moz_logins** does not delete entries from *key3.db*. Hence as long as *key3.db* is not being corrupted, after inserting the rows with previously saved values, password manager will be able to use those entries for auto login correctly (step 8).

Again in masked mode, all the root user specific credentials cached web sites are masked from further use unless and until the root user himself disable the mode. Hence to disable this mode (step 8) root user need to provide his true identity by a fool proof mechanism. Password or personal data are not entirely fool proof. One possible option is to use **biometric authentication**.

There are several techniques which are used for biometric authentication. Some well known techniques are:

- **Finger print** technology
- **Face recognition** technology
- **IRIS** technology
- **Hand geometry** technology
- **Retina geometry** technology
- **Speaker recognition** technique
- **Signature verification** technique

There are various parameters which measure the performance of any biometric authentication techniques.

Factors of evaluation are:

- **False Accept Rate(FAR)** and **False Match Rate(FMR)**
- **Equal Error Rate(EER)**
- False Reject Rate (FFR) or False Non-Match Rate (FNMR) etc.

Various studies shown that no two persons have the same fingerprints, they are unique for every individual, not even those of identical twins. Fingerprints remain constant throughout life. These properties make fingerprint a very popular biometric measurement.

In 1888 Sir Francis Galton first observed that fingerprints are rich in details which are in the form of discontinuities in ridges known as minutiae. The position of these minutiae doesn't change over time; hence minutiae matching are a good way to authenticate fingerprints.

A fingerprint recognition system consists of a fingerprint capturing device, and fingerprint matching SDK (to extract minutiae and to match minutiae). The basic idea is:

1. One or more fingerprint image is captured on the device (e.g., a personal computer system with integrated scanner).

2. This image is converted into minutiae data. Quality of these minutiae is checked to make sure that these minutiae is usable and then a template of this is stored in the Database.

3. When a user wants to log in, he must scan his finger again. It is converted again into minutiae and a second template is generated. Next, an analysis is performed by feature matching with already stored fingerprint in DB to determine if there is a match. This returns a matching score which is used to decide whether to consider this as same identity by using some threshold.

The two most important minutiae which are used to determine the match are termination and bifurcation. **Termination** is the immediate ending of the ridge and **Bifurcation** is the point on the ridge from where it is divided into two ridges.

IV. RESULTS AND DISCUSSIONS

TABLE I. SYSTEM CONFIGURATION

| Components | Specification |
| --- | --- |
| Hard Disk (Secondary storage) | 160 GB |
| RAM (Primary memory) | 2 GB |
| Processor | Intel® Core™ 2 Duo |
| Processor Speed | 2.00 GHz |

| Operating System | Ubuntu 10.04 |
|---|---|
| Compiler | Gnu Compiler Collection(gcc), javac, ECJ(Eclipse Compiler for Java) |

## A. DATABASE ACCESS

To access the SQLite Database, we have used SQLite C/C++ database programming API.

The interface to the SQLite library consists of three core functions, one opaque data structure, and some constants used as return values. The core interface is as follows:

**typedef struct sqlite sqlite ;**

**#define SQLITE OK 0        /*Successful result */**

**sqlite *sqlite_open( const char *dbname , int mode , char **errmsg ) ;**

**void sqlite_close (sqlite *db ) ;**

**int sqlite_exec (sqlite *db, char *sql,**

**int (*xCal lback ) ( void *, int , char **, char **) ,**

**void *pArg ,**

**char **errmsg**

**) ;**

The **callback** function is used to receive the results of a query.

The **sqlite_exec** function is query execution interface.

## B. RESULTS

The result of the execution of a select query is a file which contains list of URLs for which password manager stored usernames and passwords. This is used to generate a UI to select URLs to disable auto login.

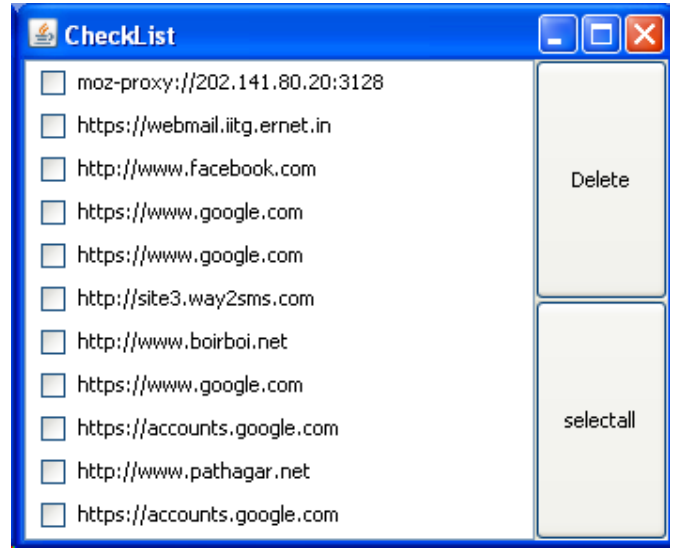

Figure 1. UI to select URLs to disable auto login.

To delete a row we simply use standard sql query for deleting. Where clause consists the hostname selected to be disabled for auto login. It will delete the corresponding row in moz_logins table. Now it will disable the auto login method. It will also save the entry in another file for future reference. Next, system will prompt for log off and log on again

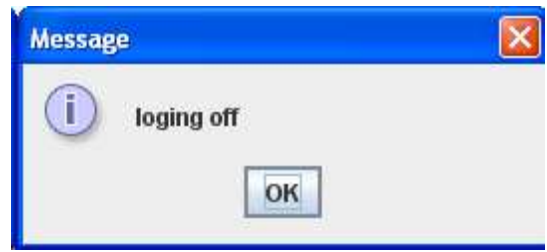

Figure 2. Prompt for login off

Now if any user tries to use any masked websites, it will ask for his own credentials.

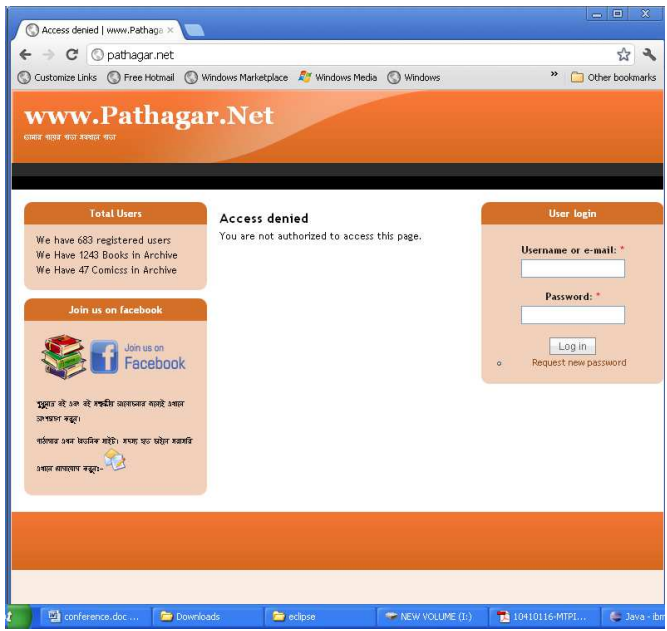

Figure 3. Prompt for user authentication for masked application

If currently logged user tries to disable the masking mode, it will ask for authentication first. As discussed earlier, the proposed method is fingerprint authentication. The Fingerprint SDK (by griaulebiometrics.com) is fingerprint recognition Software Development Kit (SDK) that allows integrating biometrics in an application. The fingerprint SDK for Java allows developing cross platform Java programs that can run even in Gnu/Linux. It can be used to develop the application which will check for authorization. The application will authorize the user to disable "**Masked Application Mode**" only if it is successful, otherwise it will prevent from do so.

Disabling is done by executing a query to insert that entry again in the table **moz_logins** using standard sql insert command with field values previously saved for corresponding hostname, it will add a row in the table **moz_logins**. Encryption and decryption keys are already present in key3.db file. We are not modifying this file and deletion from **signons.sqlite** does not delete corresponding encryption and decryption keys from this file.

Hence when root user again visits that particular website he will be able to do auto login again.

## V. CONCLUSION

In this approach sql queries are used to extract the information from the SQLite database file signons.sqlite. These queries are embedded in a C program using C SQLite programming API. Approach for biometric authentication is fingerprint matching using fingerprint scanner and fingerprint SDK. However, biometrics has also some privacy and security related problems. Main problem with biometrics is that if it is compromised there is no way to invert it. One possible approach to secure biometric authentication is use of robust hash function. The robust hash function is a one-way transformation.

REFERENCES

[1] https://developer.mozilla.org/

[2] http://kb.mozillazine.org/Password Manager

[3] http://www.symantec.com/connect/articles/password-management-concerns-ie-and-firefox

[4] http://kb.mozillazine.org/Profile folder.

[5] Biometric Authentication-Security And Usability by Vaclav Matyas and Zdenek

[6] Biometric Authentication: A Review by Debnath Bhattacharyya, Rahul Ranjan, Farkhod Alisherov A., and Minkyu Choi

[7] Biometrics: A Tool for Information Security by Jain A. K., Ross A. & Pankanti S.

[8] SQLite Embedded Database tutorial(online)

[9] The C language interface to SQLite;http://www.sqlite.org/c interface.html

[10] One-Step Query Execution Interface; http://www.sqlite.org/c3ref/exec.html

[11] A Secure Biometric Authentication Scheme Based on Robust Hashing by Yagiz Sutcu, Husrev Taha Sencar & Nasir Memon

[12] Fingerprint Identification By Kuntal Barua, Samayita Bhattacharya & Dr. Kalyani Mali

[13] Automatic Fingerprint Recognition: Minutiae based by Ispreet Singh Virk & Raman Maini.